\documentclass[11pt,a4paper]{article}
\pdfoutput=1 
\usepackage{jcappub}
\usepackage{graphicx}
\usepackage{dcolumn}
\usepackage{amsfonts}
\usepackage{amssymb}
\usepackage{bm}
\usepackage{hyperref}
\usepackage{url}
\usepackage{subfigure}
\usepackage[sort&compress]{natbib}
\usepackage{txfonts}
\usepackage{color}

\begin{document}
\newcommand{\changeR}[1]{\textcolor{red}{#1}}

\newcommand{\TBB}{{{T_{\rm BB}}}}
\newcommand{\TCMB}{{{T_{\rm CMB}}}}
\newcommand{\Te}{{{T_{\rm e}}}}
\newcommand{\Teq}{{{T^{\rm eq}_{\rm e}}}}
\newcommand{\Ti}{{{T_{\rm i}}}}
\newcommand{\nB}{{{n_{\rm B}}}}
\newcommand{\nHe}{{{n_{\rm ^4He}}}}
\newcommand{\nHet}{{{n_{\rm ^3He}}}}
\newcommand{\nHt}{{{n_{\rm { }^3H}}}}
\newcommand{\nHtw}{{{n_{\rm { }^2H}}}}
\newcommand{\nBes}{{{n_{\rm { }^7Be}}}}
\newcommand{\nLis}{{{n_{\rm { }^7Li}}}}
\newcommand{\nLisi}{{{n_{\rm { }^6Li}}}}
\newcommand{\nS}{{{n_{\rm s}}}}
\newcommand{\Teff}{{{T_{\rm eff}}}}

\newcommand{\id}{{{\rm d}}}
\newcommand{\aR}{{{a_{\rm R}}}}
\newcommand{\bR}{{{b_{\rm R}}}}
\newcommand{\neb}{{{n_{\rm eb}}}}
\newcommand{\neql}{{{n_{\rm eq}}}}
\newcommand{\kB}{{{k_{\rm B}}}}
\newcommand{\EB}{{{E_{\rm B}}}}
\newcommand{\zmin}{{{z_{\rm min}}}}
\newcommand{\zmax}{{{z_{\rm max}}}}
\newcommand{\YBEC}{{{Y_{\rm BEC}}}}
\newcommand{\YSZ}{{{Y_{\rm SZ}}}}
\newcommand{\rhob}{{{\rho_{\rm b}}}}
\newcommand{\Ne}{{{n_{\rm e}}}}
\newcommand{\sigT}{{{\sigma_{\rm T}}}}
\newcommand{\me}{{{m_{\rm e}}}}
\newcommand{\nBB}{{{n_{\rm BB}}}}

\newcommand{\kD}{{{{k_{\rm D}}}}}
\newcommand{\KC}{{{{K_{\rm C}}}}}
\newcommand{\KdC}{{{{K_{\rm dC}}}}}
\newcommand{\Kbr}{{{{K_{\rm br}}}}}
\newcommand{\zdC}{{{{z_{\rm dC}}}}}
\newcommand{\zbr}{{{{z_{\rm br}}}}}
\newcommand{\aC}{{{{a_{\rm C}}}}}
\newcommand{\adC}{{{{a_{\rm dC}}}}}
\newcommand{\abr}{{{{a_{\rm br}}}}}
\newcommand{\gdC}{{{{g_{\rm dC}}}}}
\newcommand{\gbr}{{{{g_{\rm br}}}}}
\newcommand{\gff}{{{{g_{\rm ff}}}}}
\newcommand{\xe}{{{{x_{\rm e}}}}}
\newcommand{\alphafs}{{{{\alpha_{\rm fs}}}}}
\newcommand{\YHe}{{{{Y_{\rm He}}}}}
\newcommand{\SE}{{{\dot{{\mathcal{E}}}}}}
\newcommand{\SQ}{{{{{\mathcal{E}}}}}}
\newcommand{\SN}{{\dot{\mathcal{N}}}}
\newcommand{\Sn}{{{\mathcal{N}}}}
\newcommand{\muc}{{{{\mu_{\rm c}}}}}
\newcommand{\xc}{{{{x_{\rm c}}}}}
\newcommand{\xH}{{{{x_{\rm H}}}}}
\newcommand{\mT}{{{{\mathcal{T}}}}}
\newcommand{\Ob}{{{{\Omega_{\rm b}}}}}
\newcommand{\Or}{{{{\Omega_{\rm r}}}}}
\newcommand{\Odm}{{{{\Omega_{\rm dm}}}}}
\newcommand{\mdm}{{{{m_{\rm WIMP}}}}}
\newcommand{\Acmb}{{{{A_{\rm CMB}}}}}
\newcommand{\Ayco}{{{{A_{\rm y/CO}}}}}
\newcommand{\Ad}{{{{A_{\rm dust}}}}}
\newcommand{\Td}{{{{T_{\rm dust}}}}}
\newcommand{\betad}{{{{\beta_{\rm dust}}}}}
\newcommand{\fyco}{{{{f_{\nu}^{\rm y/CO}}}}}
\newcommand{\nudo}{{{{\nu_{0}^{\rm dust}}}}}
\newcommand{\fnud}{{{{f_{\nu}^{\rm dust}}}}}
\newcommand{\npl}{n_{\rm pl}}
\newcommand{\nmu}{n_{\mu}}
\newcommand{\Tref}{T_{\rm ref}}
\newcommand{\xref}{x_{\rm ref}}
\newcommand{\Og}{{{{\Omega_{\rm r}}}}}
\newcommand{\bx}{\mathbf{x}}
\newcommand{\bk}{\mathbf{k}}
\newcommand{\bkp}{\mathbf{{k}'}}
\newcommand{\hbn}{\mathbf{\hat{n}}}
\newcommand{\hbk}{\mathbf{\hat{k}}}
\newcommand{\hbkp}{\mathbf{\hat{k}'}}
\newcommand{\mR}{\mathcal{R}}
\newcommand{\mP}{\mathcal{P}}
\newcommand{\mG}{\mathcal{G}}
\newcommand{\Rnu}{R_{\nu}}
\newcommand{\rs}{r_{\rm s}}
\newcommand{\fnl}{f_{\rm NL}}
\newcommand{\gnl}{g_{\rm NL}}
\newcommand{\taunl}{\tau_{\rm NL}}
\newcommand{\zpl}{z_{\rm pl}}
\newcommand{\etapl}{\eta_{\rm pl}}

\title{Constraints on $\mu$-distortion fluctuations and primordial
  non-Gaussianity from Planck data}

\author[a]{Rishi Khatri,}
\author[a,b,c]{Rashid Sunyaev}

\affiliation[a]{ Max Planck Institut f\"{u}r Astrophysik\\, Karl-Schwarzschild-Str. 1
  85741, Garching, Germany }
\affiliation[b]{Space Research Institute, Russian Academy of Sciences, Profsoyuznaya
 84/32, 117997 Moscow, Russia}
\affiliation[c]{Institute for Advanced Study, Einstein Drive, Princeton, New Jersey 08540, USA}
\date{\today}
\emailAdd{khatri@mpa-garching.mpg.de}
\abstract
{We use the Planck HFI channel maps to make an all sky map of
  $\mu$-distortion fluctuations. Our $\mu$-type distortion map is dominated
  by
   the $y$-type distortion contamination from the hot gas in the
low redshift Universe and we can thus only place upper limits on the
$\mu$-type distortion fluctuations. For the
amplitude of $\mu$-type distortions on $10'$ scales we get the limit on root mean square
(rms) value $\mu_{\rm
  rms}^{10'}< 6.4\times 10^{-6}$, a limit 14 times stronger than the
  COBE-FIRAS ($95\%$ confidence) limit on the mean of $\langle \mu \rangle<90\times
  10^{-6}$. Using our maps we also place strong upper limits on the auto angular power spectrum of $\mu$,
  $C_{\ell}^{\mu\mu}$ and the cross angular power spectrum of $\mu$ with the 
 CMB temperature anisotropies, $C_{\ell}^{\mu T}$. The strongest
 observational limits are
 on the largest scales,
 $\ell(\ell+1)/(2\pi)C_{\ell}^{\mu\mu}|_{\ell=2-26}<(2.3\pm 1.0)\times
 10^{-12}$ and $\ell(\ell+1)/(2\pi)C_{\ell}^{\mu T}|_{\ell=2-26}<(2.6\pm
 2.6)\times 10^{-12}~{\rm K}$. Our observational limits can be used to
 constrain new physics which can create spatially varying energy release in
 the early Universe between redshifts $5\times 10^4\lesssim z\lesssim
 2\times 10^6$. We specifically apply our observational results  to
 constrain the primordial
  non-Gaussianity of the local type, when the source of $\mu$-distortion is
  Silk damping, for very squeezed configurations with the wavenumber for the
  short wavelength mode  $46 \lesssim  k_{\rm S} \lesssim 10^4 ~{\rm
  Mpc^{-1}}$ and for the long wavelength mode $k_{\rm L}\approx 10^{-3} ~{\rm
  Mpc^{-1}}$.  Our limits on the primordial non-Gaussianity parameters are
  $f_{\rm NL}<10^5, \taunl<1.4\times 10^{11}$ for  $k_{\rm S}/k_{\rm L}\approx
  5\times 10^4- 10^7$. We also give a new derivation of the evolution of
the  $\mu$-distortion fluctuations through the $y$-distortion era and the
recombination epoch until today resulting in very simple expressions for the
cross and auto power spectra in the squeezed limit. We also introduce
mixing of Bose-Einstein spectra due to Silk damping and $y^{\rm BE}$-type
distortions. The $\mu$-type distortion map and masks are now publicly available.
}

\maketitle
\flushbottom
\section{Introduction}
Complete thermodynamic equilibrium between the photons and the
electromagnetic plasma exists only in the very early Universe in the
standard cosmological model. The memory of any event which injects energy
into the electromagnetic plasma is quickly erased under such circumstances, apart from the change in
the temperature of the plasma and the photons. This fundamental limit on how far
and how well 
we can probe the early Universe using the CMB photons is much further than
the last scattering surface at $z\sim 1100$ \cite{zks68,peebles68,sz1970c} that we
observe with the CMB anisotropy experiments such as COBE-DMR (Cosmic
Background Explorer - Differential Microwave Radiometer) \cite{cobedmr}, WMAP (Wilkinson
Microwave Anisotropy Probe) \cite{wmap}, Planck \cite{planck},
SPT (South Pole Telescope) \cite{sptdamping}, ACT (Atacama Cosmology
Telescope) \cite{actdamping} and many others.\footnote{See \url{http://http://lambda.gsfc.nasa.gov/links/experimental_sites.cfm} for a
  complete list of CMB experiments} 

An important
landmark across which the properties of the Universe change in a
significant way in a short time is the blackbody surface
\cite{sz1970,ks2012} at $z\approx 2\times 10^6$. At times earlier than this
landmark,
$z\gtrsim 2\times 10^6$, the photon creation at low frequencies ($x\equiv
\nu/T \ll 1$, where $\nu$ is the photon  frequency, $T$ is the temperature
of the electromagnetic plasma, $x$ is the dimensionless frequency and we
are using natural units with Planck's constant ($h$) and Boltzmann constant
($\kB$) 
equal to unity.) combined with the redistribution of photons over the whole
spectrum by Compton scattering maintains a Planck spectrum for the photons
\cite{sz1970}. The dominant process for the photon creation is the double
Compton scattering for a low baryon density Universe such as ours
\cite{dd1982} with a small contribution from bremsstrahlung. At lower
redshifts, $z\lesssim 2\times 10^6$, the photon creation by double Compton
and bremsstrahlung becomes inefficient, making the photon number a conserved
quantity. Any injection of energy is still redistributed by Compton
scattering but the inability to create and destroy photons means that the
equilibrium spectrum is not the Planck spectrum but the Bose-Einstein
spectrum (see e.g. \cite{llstats}) with a dimensionless chemical potential
($\mu\equiv -\mu_E/T$, where $\mu_E$ is the thermodynamic chemical
potential) that does not
vanish in general,
\begin{align}
n_{\rm BE}(x)&=\frac{1}{e^{x+\mu}-1}.
\end{align}
{The Bose-Einstein spectrum is a stationary solution of the Kompaneets
equation \cite{k1956,sz1970}.}

For observational purpose and also for comparison with the $y$-type
distortions \cite{zs1969} it is convenient to define the dimensionless
frequency not with respect to the temperature of the Bose-Einstein spectrum
but with respect to the temperature of a Planck spectrum which has the same
number density of photons as the Bose-Einstein spectrum, $N_{\rm
  BE}(T,\mu)\equiv N_{\rm Pl}(\Tref) \Rightarrow (T-\Tref)/\Tref\approx 0.456\mu$,$\xref\equiv \nu/\Tref$. In terms of
this reference temperature  we have
\begin{align}
n_{\rm BE}(\xref)&\equiv\frac{1}{e^{\xref\Tref/T+\mu}-1}\nonumber\\
&\overset{\mu\ll 1}\approx \frac{1}{e^{\xref}-1}+\frac{\mu
  e^\xref}{\left(e^{\xref}-1\right)^2}\left(\frac{\xref}{2.19}-1\right)\label{Eq:mu}\nonumber\\
&\equiv \npl+\mu\nmu,
\end{align}
where the first term is the Planck spectrum and the second term defines the
$\mu$-type distortion in the limit of small distortions \cite{sz1970}. Note that at
linear order, for small $\mu$, we have $\mu\equiv -\mu_E/T \approx
-\mu_E/\Tref$. We should emphasize that this re-definition is just for
convenience of visualization and does not affect the analysis of data. This
is because 
the monopole temperature of the CMB is also uncertain, at the sensitivity we will be
working, and we must fit for the reference temperature simultaneously giving
us the freedom to choose the reference temperature. We will use the definition
$n_{\mu}$ from Eq. \ref{Eq:mu} throughout the rest of the paper with
$T=2.725~{\rm K}$ as the reference temperature  dropping the subscript
${}_{\rm ref}$. 

The $\mu$-type distortions thus probe the era that lies behind the last scattering
surface and are  sensitive to any process that injects energy into the
electromagnetic plasma \cite[see e.g.][]{cs2011,sk2013} between redshifts
$5\times 10^4 \lesssim z \lesssim 2\times 10^6$ \cite{ks2012b}. Most
processes, owing to the statistical homogeneity of the Universe, give
distortions which are identical in different regions of the
Universe. Planck is insensitive to the absolute brightness of the sky and
therefore cannot detect the constant or invariant component of the
$\mu$-type distortions. One of the exceptions is the dissipation of primordial
perturbations or sound waves in the electromagnetic plasma (Silk damping)
\cite{silk,Peebles1970,kaiser} before recombination. The primordial
perturbations  excite {standing} sound waves in the tightly coupled
baryon-photon fluid as they enter the horizon
\cite{lifshitz,sz1970c,Peebles1970}. The mean free path of the photons for
Thomson scattering on free electrons is very small during this time, they can however still
diffuse to much larger scales doing a random walk in the sea of
electrons. The (comoving) diffusion scale ($\lambda_{\rm D}$) and
wavenumber ($\kD$) are given during the radiation dominated era by (including both thermal
conductivity and radiative viscosity) \cite{silk,Peebles1970,kaiser}
\begin{align}
\lambda_{\rm D}=\frac{2\pi}{k_{\rm D}}&=
\left[\int_0^{\eta}\id \eta
\frac{2\pi c(1+z)}{6(1+R)\Ne \sigT
}\left(\frac{R^2}{1+R}+\frac{16}{15}\right)\right]^{1/2}\nonumber\\
&\approx \left(\frac{16\pi c}{135\Og^{1/2}H(0)\Ne(0)\sigT(1+z)^3}\right)^{1/2},\label{Eq:kd}
\end{align}
where $R\equiv \frac{3\rho_b}{4 \rho_{\gamma}}$, $\rho_b$ is the baryon
  energy density, $\rho_{\gamma}$ is the photon energy density,
 $\eta$ is conformal time, $\sigT$ is the Thomson cross section, $\Ne(\eta)$ is
 the electron number density, $H(\eta)$ is the Hubble parameter, $\Or$ is
 the radiation energy density in units of critical density today,
 $\eta=0$ is present time and   
 $c$ is the speed of light.  The diffusion or dissipation scale at
 $z=5\times 10^4$ is $k_{\rm D}\approx 46~{\rm Mpc^{-1}}, \lambda_{\rm
   D}\approx 0.14~{\rm Mpc}$ and at $z=2\times 10^6$ it is  $k_{\rm D}\approx
 1.1\times 10^4~{\rm Mpc^{-1}}, \lambda_{\rm
   D}\approx 0.55~{\rm kpc}$. The photon diffusion erases the perturbations
 on the diffusion scales, moving the energy in the perturbations or sound
 waves into the
 local CMB monopole spectrum, giving a chemical potential to the
 spectrum \cite{sz1970,daly1991,hss94,cks2012,ksc2012b}. If the primordial curvature
 perturbations are Gaussian, there is same power on the small scales
 $46\lesssim k_{\rm D}\lesssim 10^4~{\rm Mpc^{-1}}$ in different regions of
 the Universe, apart from the cosmic variance $\propto N_{\rm
   modes}^{1/2}=  (k_{\rm L}/k_D)^{3/2}$, where
 $k_{\rm L} \lesssim 0.01~ {\rm Mpc^{-1}}$ is the separation of different regions in the
 Universe over which we wish to study the fluctuations. If the primordial
 non-Gaussianity of local type is present on these scales \cite{pajer2012},
 the small scale power is correlated with the large scale fluctuations and
 the $\mu$-distortion has large scale anisotropy on the sky with amplitude dependent on
 the amplitude of the primordial non-Gaussianity. The fluctuations in the
 $\mu$-type distortions, and hence the primordial non-Gaussianity on these
 scales, is already accessible with Planck, which has much  higher sensitivity
 compared to the COBE-FIRAS {for a fluctuating signal} \cite{cobe}.

We will use the component separation method {\sc LIL} (Linearized Iterative
Least-squares) \cite{k2014} to separate the $\mu$-distortion component from
Planck data. The small number of channels and similarity of the
$\mu$-distortion spectrum with the $y$-type spectrum means that our map
would be dominated by contamination from the low redshift $y$-type
distortions and we will only be able to put upper limits on the
$\mu$-distortion fluctuations and primordial non-Gaussianity. In principle,
there is additional information in the intermediate-type distortions
\cite{ks2012b,ks2013b,chlubajeong2013,clesse2014} which is however
inaccessible with Planck owing to the limited number of frequency channels.
The maps and masks, {created by one of us (RK),} are made
publicly available at \url{http://www.mpa-garching.mpg.de/~khatri/muresults/}.

\section{$\mu$-type distortions in Planck channels}
We will be using the 4 HFI channels of the Planck (100, 143, 217, 353
GHz). Each of these channels has finite bandwidth and we must integrate the
 intensity spectrum of $\mu$-type distortions ($2 h \nu^3/c^2n_{\mu}$) over
 each spectral band. In addition it is also convenient to convert the
 intensity into thermodynamic CMB units defined by relating the change in
 intensity in the spectral band, $\Delta I_{\nu}$, to the change in the
 intensity $I_{\rm pl}(T)$ of the Planck spectrum for a change in temperature $\Delta T_{\rm K_{CMB}}$,
\begin{align}
\Delta I_{\nu}&= \Delta T_{\rm K_{CMB}}\frac{\partial I_{\rm pl}}{\partial
  T}\nonumber\\
&=\Delta T_{\rm K_{CMB}}\frac{2 h \nu^3}{T c^2}\frac{xe^x}{(e^x-1)^2}
\end{align}
The units of the maps released by Planck are ${\rm K_{CMB}}$
  for the 4 HFI channels we will be using. To convert any other spectrum,
  $n(x)$,  to
  the CMB temperature units we should integrate both the intensity of the
  spectrum of interest and differential Planck spectrum over the
  transmission profile of the channels $w(\nu)$ \cite{planckhfi} which are
  available on the Planck legacy archive,
\begin{align}
\Delta T_{\rm K_{CMB}}=\frac{\int w(\nu) 2h\nu^3/c^2 n(\nu) \id \nu}{\int w(\nu)\partial I_{\rm pl}/\partial T \id
  \nu}.
\end{align}
  
For the $\mu$ distortion $n(x)=n_{\mu}$, and similarly for the
$y$-type distortion we substitute $n(x)=n_{\rm y}$ where $n_{\rm y}$ is the $y$-type
spectrum \cite{zs1969}
\begin{align}
n_{\rm y}=\frac{xe^x}{(e^x-1)^2}\left[x\left(\frac{e^x+1}{e^x-1}\right)-4\right]
\end{align}

\begin{figure}
\resizebox{\hsize}{!}{\includegraphics{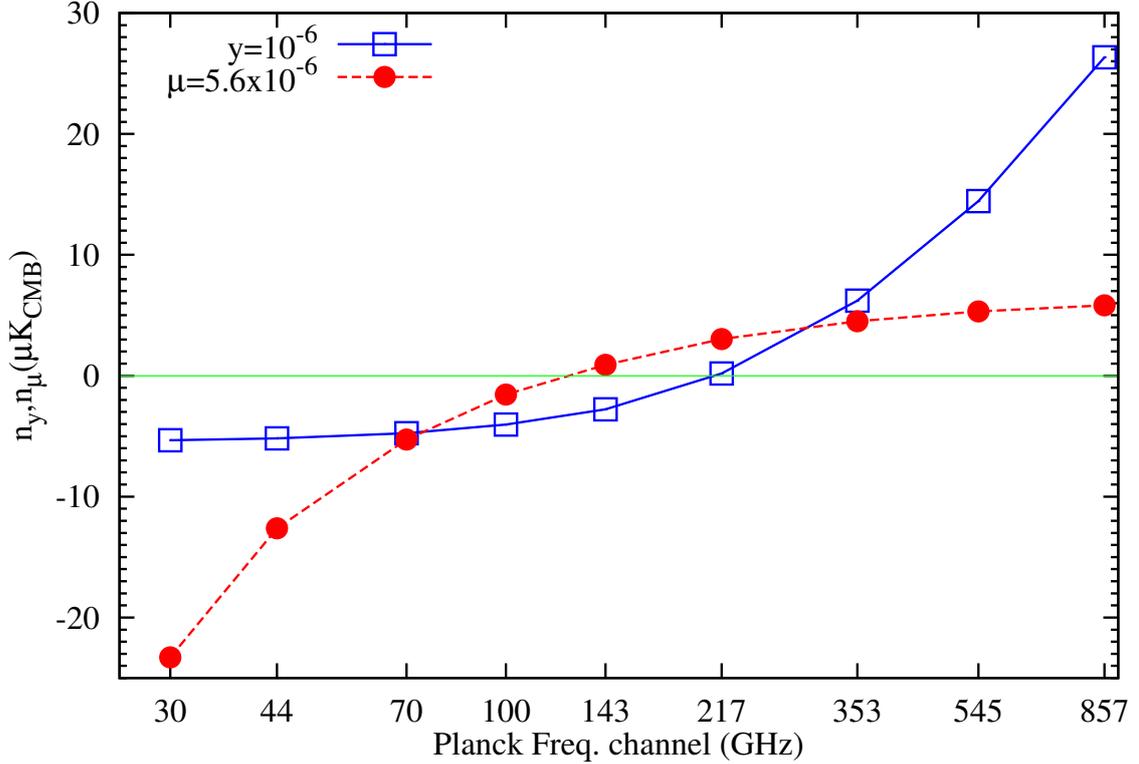}}
\caption{\label{Fig:ymuspec}
The spectra of $y$-type and $\mu$-type distortions in the $\mu {\rm
  K_{CMB}}$ units as seen by Planck channels. {We will only use the 4 HFI
channels, 100 GHz to 353 GHz, in our analysis.}
}
\end{figure}

We give the conversion factors calculated for all Planck channels for both $\mu$ and
$y$-type distortions in Table
\ref{tbl:conv} and the spectra are plotted in Fig. \ref{Fig:ymuspec}. Our
$y$ to ${\rm K_{CMB}}$ conversion factors match those released by Planck
\cite{planckhfi} with a small difference because of our using the band
transmission profile from the second
data release. We will exploit the small difference in the two spectra to
mask out as much of the $y$-distortion signal from galaxy clusters and
groups as possible allowing us to put much stronger constraints on the
$\mu$-type distortions compared to the constraints we would expect in the absence of such a mask.
\begin{table}
\begin{tabular}{|c|c|c|c|c|c|c|c|c|c|}
\hline
 Channel-GHz &30 &44&70&100&143&217&353&545&857\\
\hline
$y$ to ${\rm K_{CMB}}$ &-5.3337&-5.1752&-4.7503&-4.0309&-2.7823&0.1941&6.2065&14.453&26.332\\
$\mu$ to ${\rm K_{CMB}}$&-4.1611&-2.2528&-0.9414&-0.2767&0.1606&0.5411&0.8053&0.9484&1.0413\\
\hline
\end{tabular}
\caption{\label{tbl:conv}Conversion factors from $y$ and $\mu$ to ${\rm
    K_{CMB}}$. We have checked that the small difference in the $y$
  conversion factors with respect to the \cite{planckhfi} are due to the
  change in the transmission profile between the first and the second
  Planck data releases. We have used the later release to calculate the
  conversion factors. All values are for the full channels.} 
\end{table}

\section{A $\mu$-type distortion map from Planck HFI data}
The component separation method, {\sc LIL}, is described in \cite{k2014}
and its application to separate the $y$-type distortion from CMB and
foregrounds is described in \cite{k2015}. The algorithm to create the
$\mu$-type distortion map is identical to that described in \cite{k2015}
with the $y$-type distortion spectrum replaced by the $\mu$-distortion
spectrum given in Table \ref{tbl:conv}. We will only sketch the main
features of the algorithm below and refer the reader to \cite{k2014,k2015}
for details of the implementation.

{\sc LIL} is a parameter fitting algorithm and we will fit a parametric
model consisting of CMB + dust + $\mu$-distortion or CMB + dust +
$y$-distortion to the 4 lowest Planck HFI channels at each pixel. The spectra of CMB and
distortions are of course fixed and the only free parameters are the
amplitudes. For the dust we fit a gray body dust model with temperature
fixed to 18 K but allowing the spectral index $\beta_{\rm dust}$ to vary between the limited
range of $2<\beta_{\rm dust}<3$ away from the galactic plane \cite[see][for
exact definitions]{k2014}. The constraint on the $\beta_{\rm dust}$ means
that the effective degrees of freedom is non-zero even though we fit 4
parameters to 4 data points. This is because away from the galactic plane,
the main region of interest for us, the signal in 217 GHz channel is too
small to constrain $\beta_{\rm dust}$ and we rarely reach a global minimum
in $\chi^2$ in this parameter direction owing to the hard constraints  imposed by us.

An advantage of the parameter fitting approach over the internal linear
combination based methods \cite{milca,nilc,planckymap} is that we get a
quantitative goodness of fit estimate which tells us if the model being fit
is the correct one. In particular, if we fit a $\mu$-type distortion
spectrum to the pixels corresponding to clusters and groups of galaxies, we
will get a $\chi^2$ which would be worse compared to the correct model with
$y$-type distortion. We use this difference in $\chi^2$ between the two
models to mask out clusters and groups according to the following algorithm

\begin{enumerate}
\item We mask a pixel if $\chi^2_y-\chi^2_{\mu}<-0.5$
\item We add the requirement that minimum hole size should be $10$ pixels
  at HEALPix \cite{healpix} resolution of nside=2048 i.e. if less than 10
  contiguous pixels are getting masked then these pixels are unmasked.
\item We also mask all pixels with negative $y$-type distortion and
  $\chi^2_y>3.8$. This is signature of radio point sources.
\item We augment the resulting mask by additional 3 pixels
\end{enumerate}

\begin{figure}
\resizebox{\hsize}{!}{\includegraphics{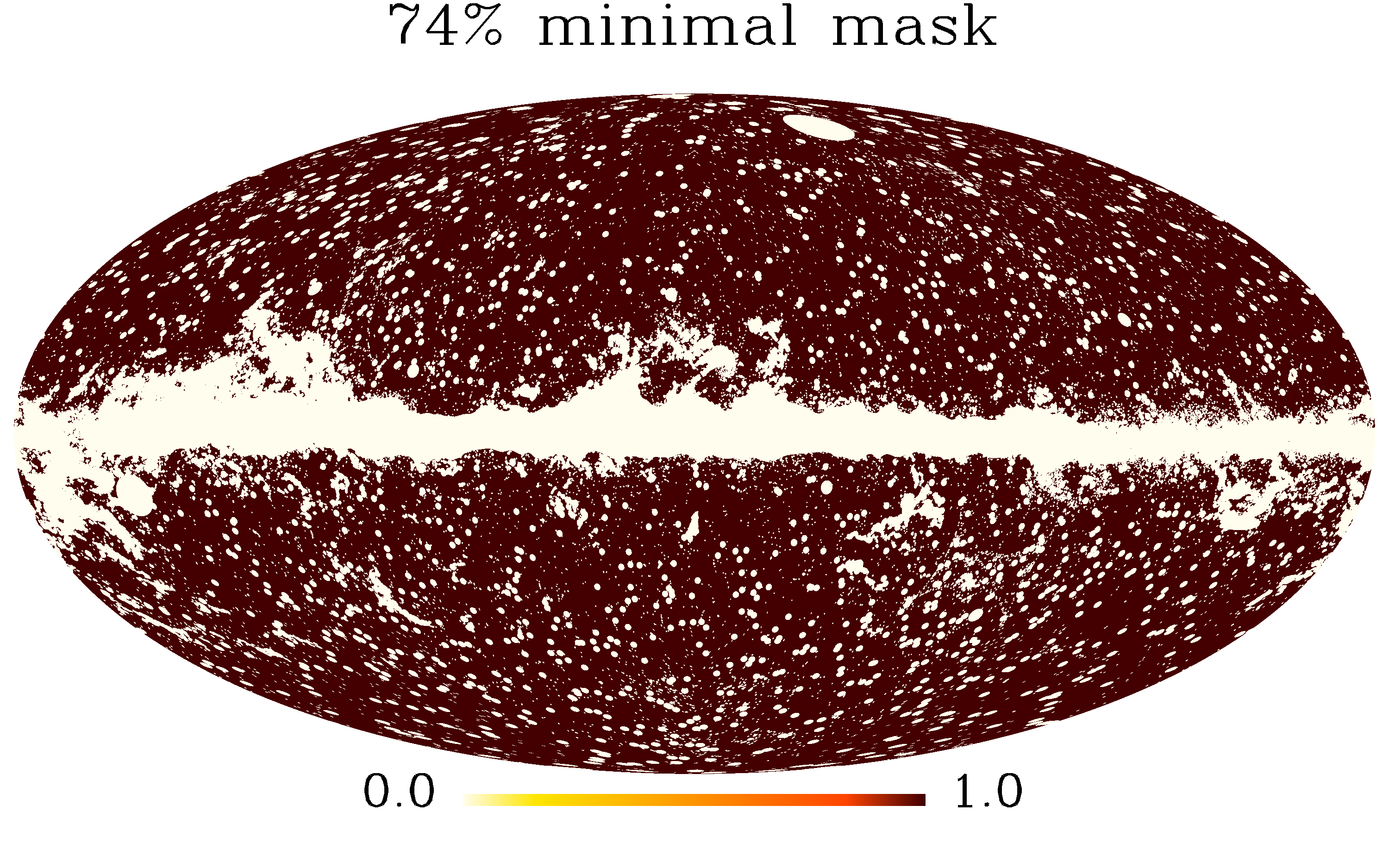}}
\caption{\label{Fig:maskmu}Minimal mask for $\mu$-distortion analysis
  masking clusters and groups of galaxies, point sources and galactic plane.
}
\end{figure}

This simple algorithm masks most of the clusters and groups that are
detected in our $y$-distortion map. In addition we explicitly mask
$4.5^{\circ}$ radius regions around Virgo and Perseus, $1.5^{\circ}$ radius
regions around all clusters with $S/N>30$ in the second Planck cluster
catalog \cite{planckclusters2015} and $0.75^{\circ}$ radius regions around
all sources in the second Planck cluster
catalog. We finally add the $86\%$ CO/point source/galactic mask that was calculated in
\cite{k2015} and is publicly available at
\url{http://www.mpa-garching.mpg.de/~khatri/szresults/}. Our final minimal
mask for the $\mu$-distortion analysis, masking
$25.8\%$ of the sky, is shown in Fig. \ref{Fig:maskmu}. We will augment
this minimal mask using thresholds on the 545 GHz channel map of  Planck to
select successively cleaner portions of the sky to test the robustness of
our results to the dust contamination.

\begin{figure}
\resizebox{\hsize}{!}{\includegraphics{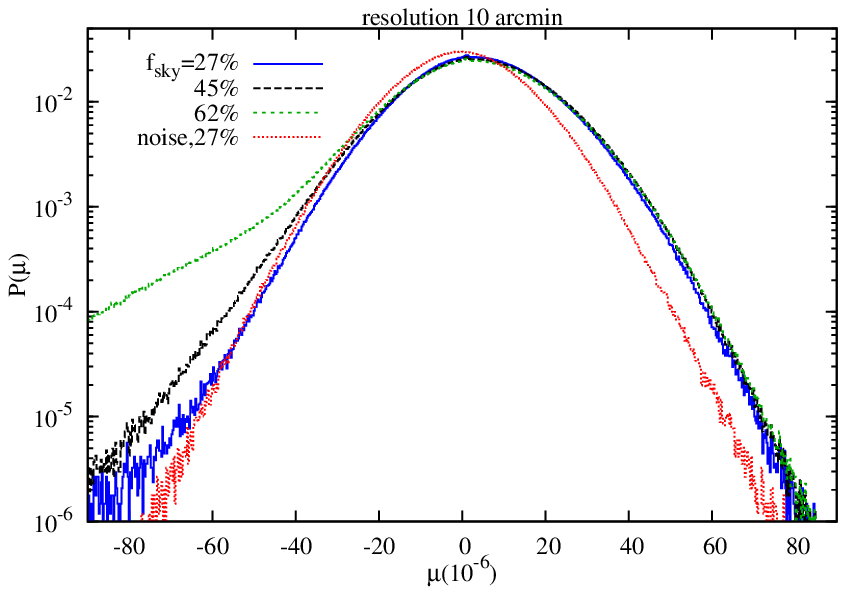}}
\caption{\label{Fig:mupdf}Probability distribution of $\mu$-distortion, $
  P(\mu)$, for different sky fractions, $f_{\rm sky}$ along with the noise for 
$f_{\rm sky}=27\%$ at $10'$ resolution calculated from the half-ring
difference maps.}
\end{figure}

\begin{figure}
\resizebox{\hsize}{!}{\includegraphics{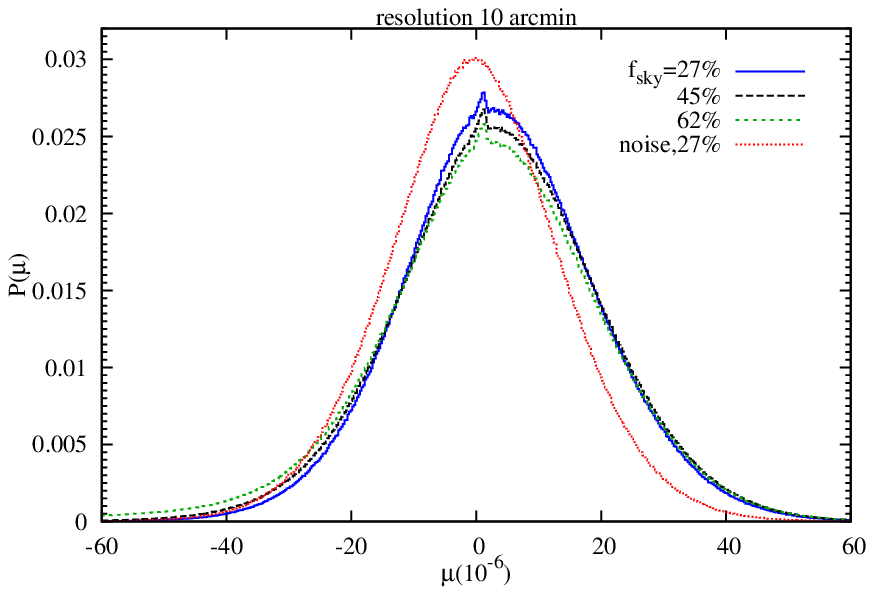}}
\caption{\label{Fig:mupdflin}Same as Fig. \ref{Fig:mupdf} but with linear
  scale on the $y$ axes.}
\end{figure}

The probability distribution function (PDF) of the $\mu$-distortion map is shown
in Fig. \ref{Fig:mupdf} for different sky fractions $f_{\rm sky}$.
The PDF is dominated by noise and residual $y$-distortion contamination at
all sky fractions. For the $62\%$ sky fraction there is also a small amount of
contamination from the dust (since the masks were designed using the 545
GHz map which is dominated by dust) which becomes negligible at $f_{\rm
  sky}=45\%$ and goes away completely at $f_{\rm
  sky}=27\%$. The contribution of the tails to the map variance and power
spectrum is negligible. This is apparent immediately if we look at the same
plot drawn on linear scale in Fig. \ref{Fig:mupdflin}. {The sharp
  peak near the small values of $\mu$ is probably due to a small number of
  clean low noise pixels where the noise dominates over the contamination.} The skewness towards
positive values because of the residual $y$-distortion contamination is
also apparent.

\section{Constraints on the $\mu$-distortion fluctuations}
\subsection{Amplitude of fluctuations}
The simplest constraint we can get on the $\mu$-distortion fluctuations is
using the total power or variance of the map. We should in particular
subtract the contribution of the noise.  We  make half ring
$\mu$-maps from the 
Planck half-ring channel maps and calculate the half ring half difference
(HRHD) $\mu$-map which gives us the white noise
estimate  in our full channel $\mu$-map. The total map variance
$\sigma_{\rm map}^2$ is sum of noise contribution, $\sigma_{\rm noise}$ and
signal $\mu_{\rm rms}^2$, assuming that the noise is uncorrelated with the signal,
\begin{align}
\sigma_{\rm map}^2=\mu_{\rm rms}^2+\sigma_{\rm noise}^2,
\end{align}
 where $\mu_{\rm rms}$ is the root mean squared value of the signal. We are
 in particular interested in the central variance, $(\mu_{\rm
   rms}^{c})^2\equiv \mu^2_{\rm rms}-\langle \mu\rangle^2$, where the angular brackets
 indicate the ensemble average. The rms and mean values of $\mu$ from our
 map for different sky fractions are given in Table \ref{tbl:mu}. Since our
 algorithm is non-linear, the half-ring half difference (HRHD) map underestimates the
 noise in the full channel map. A better, unbiased, estimate of the signal
 variance would be if we average the product of the two half-ring maps.
 The half-ring maps have uncorrelated white noise which automatically
 cancels in the  cross-variance, 
\begin{align}
(\sigma_{\rm map}^X)^2&\equiv \langle \mu^{\rm hr1}(p)\mu^{\rm hr2}(p)\rangle
\nonumber\\
&=(\mu_{\rm rms}^X)^2,
\end{align}
where hr1 and hr2 refer to the half-ring maps, the ensemble average is over
all pixels $p$, and $X$ refers to the cross product of two maps.
The central variance is then given by,  $(\mu_{\rm
   rms}^{cX})^2\equiv (\mu_{\rm rms}^X)^2-\langle \mu\rangle^2$. We will
 use $\mu_{\rm
   rms}^{cX}$ as our upper limit.

 For our
 smallest mask $f_{\rm sky}=70.7\%$, the results are clearly dominated by
 the galactic contamination. The results however vary less rapidly from
 $f_{\rm sky}=62.3\%$ onward and seem to be converging with decreasing sky
 fractions. Assuming that all the signal is coming from the $y$-distortion
 contamination we get an upper limit of $\mu_{\rm
   rms}^{\rm central}<6.1\times 10^{-6}$. Note that since we are averaging over
 millions of pixels, even for the smallest sky fraction, the statistical
 error is negligible even after taking into account the fact that the noise
 in the neighboring pixels is correlated.

\begin{table}
\begin{tabular}{|c|c|c|c|c|c|c|}
\hline
  $f_{\rm sky}$  &$\sigma_{\rm map}$ &$\sigma_{\rm noise}$&$\langle{\mu}\rangle$&$\mu_{\rm rms}$&$\mu_{\rm rms}^c$ &$\mu_{\rm rms}^{cX}$\\
&$(\times 10^{-6})$&$(\times 10^{-6})$&$(\times 10^{-6})$&$(\times 10^{-6})$&$(\times 10^{-6})$&$(\times 10^{-6})$\\
\hline
$70.7\%$&24.23&18.00&-0.40&16.21&16.21&9.81\\
$62.3\%$&18.67&15.69&1.87&10.13&9.95&7.35\\
$54.0\%$&17.40&15.15&2.85&8.57&8.08&6.77\\
$45.1\%$&16.80&14.77&3.48&8.01&7.21&6.51\\
$36.6\%$&16.43&14.48&3.77&7.77&6.79&6.37\\
$26.9\%$&15.95&14.14&3.70&7.38&6.39&6.10\\
\hline
\end{tabular}
\caption{\label{tbl:mu}Moments and other properties of the $\mu$-distortion map PDF for different
  sky fractions. The last column ($\mu_{\rm rms}^{cX}$) is the  limit on the
  $\mu$-distortion fluctuations calculated from cross-variance of half-ring
  maps while $\mu_{\rm rms}^{c}$ is calculated by subtracting noise
  estimate from half-ring half difference (HRHD) maps from the full channel map. The
  later slightly overestimates the signal variance.
}
\end{table}

\subsection{Auto and cross power spectrum}
\begin{figure}
\resizebox{\hsize}{!}{\includegraphics{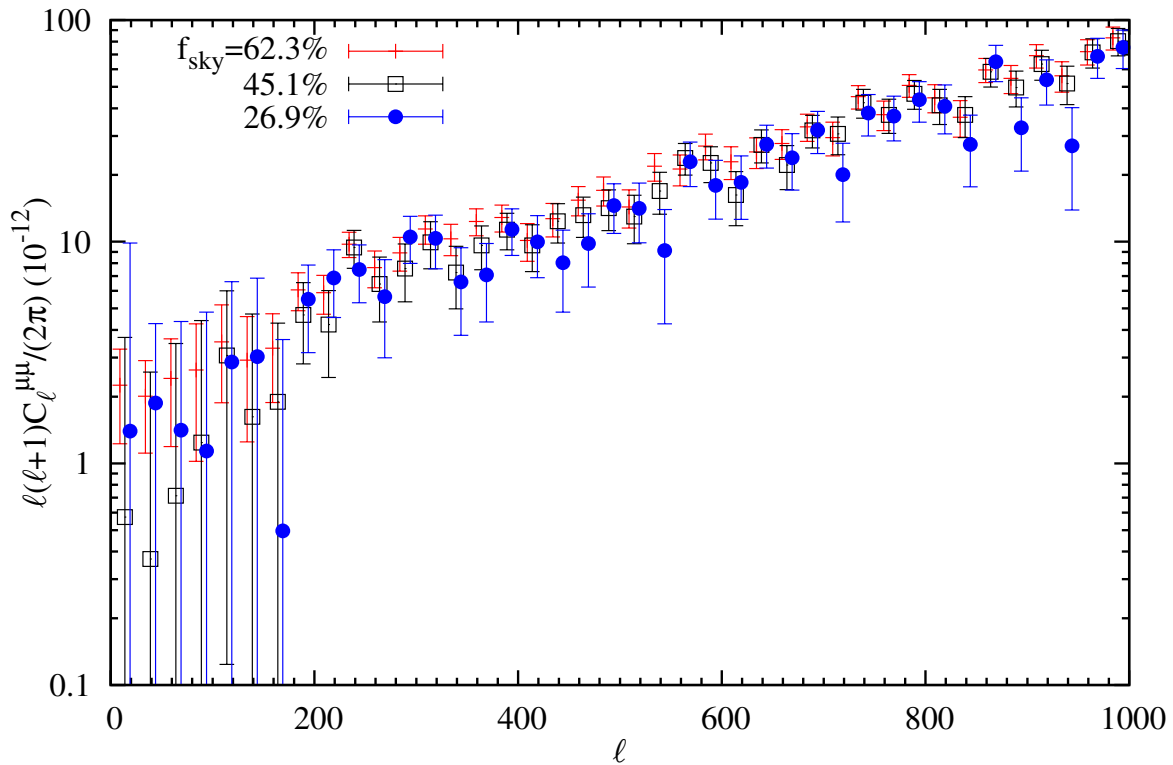}}
\caption{\label{Fig:clmm}Auto power spectrum of $\mu$-type distortion
  fluctuations. The effect of the effective beam of $\mu$-map (10' FWHM)
  and mask has been corrected for. The points for different masks have been
offset on the x-axes from the center of the $\Delta \ell=100$ bins to make them distinguishable.}
\end{figure}

We can also calculate the angular power spectrum of our map,
$C_{\ell}^{\mu\mu}$ and the cross power spectrum with the CMB temperature map, $C_{\ell}^{\mu T}$.  For calculating the power
spectra we use publicly available PolSpice code \cite{ps1,ps2} which
deconvolves the effects of masks \cite{hivon2002} in correlation space
making it very fast. The code also calculates the covariance matrix
\cite{e2004,cc2005,xspect} that we use to estimate the errors. We use large
bins of $\Delta \ell=25$ to mitigate to some extent the effects of mode
coupling due to our very complex mask. We augment our minimal mask using
the 545 GHz channel map smoothed to 2 degree FWHM beam to create the masks with
successively cleaner regions of the sky. These masks were  used in Table
\ref{tbl:mu}. For calculating the $C_{\ell}$ we apodize the mask with a
Gaussian function in pixel space by replacing the $1s$ in the mask by
$1-\exp\left[-9\theta^2/2(\theta_{\rm ap})^2\right]$ for
$\theta<\theta_{\rm ap}=30'$, where $\theta$ is the distance of the pixel
from the edge of the mask. To get an estimate of the power spectrum
unbiased by noise we will use the cross-power spectrum of the half-ring maps
automatically subtracting the uncorrelated noise. The auto power spectrum
calculated in this way is shown in
Fig. \ref{Fig:clmm} for several sky fractions corrected for the $10'$ beam
of our maps in addition to the effect of the mask.
There is a small decrease in the power on small scales when the sky
fraction is reduced signifying the level of galactic contamination. On
large scales, where we have the best constraints,  the power
spectra for different masks are consistent with each other (taking into
account the error bars) and we will use
the $f_{\rm sky}=62.3\%$ power spectrum on large scales to put limits on
the $\mu$-distortion fluctuations and primordial non-Gaussianity. Thus we
have the limit on the power spectrum  in the $\ell =2-26$ bin,
\begin{align}
\ell(\ell+1)C_{\ell}^{\mu\mu}/(2\pi)|_{\ell=2-26}<(2.3\pm
1.0) \times 10^{-12}
\end{align}

\begin{figure}
\resizebox{\hsize}{!}{\includegraphics{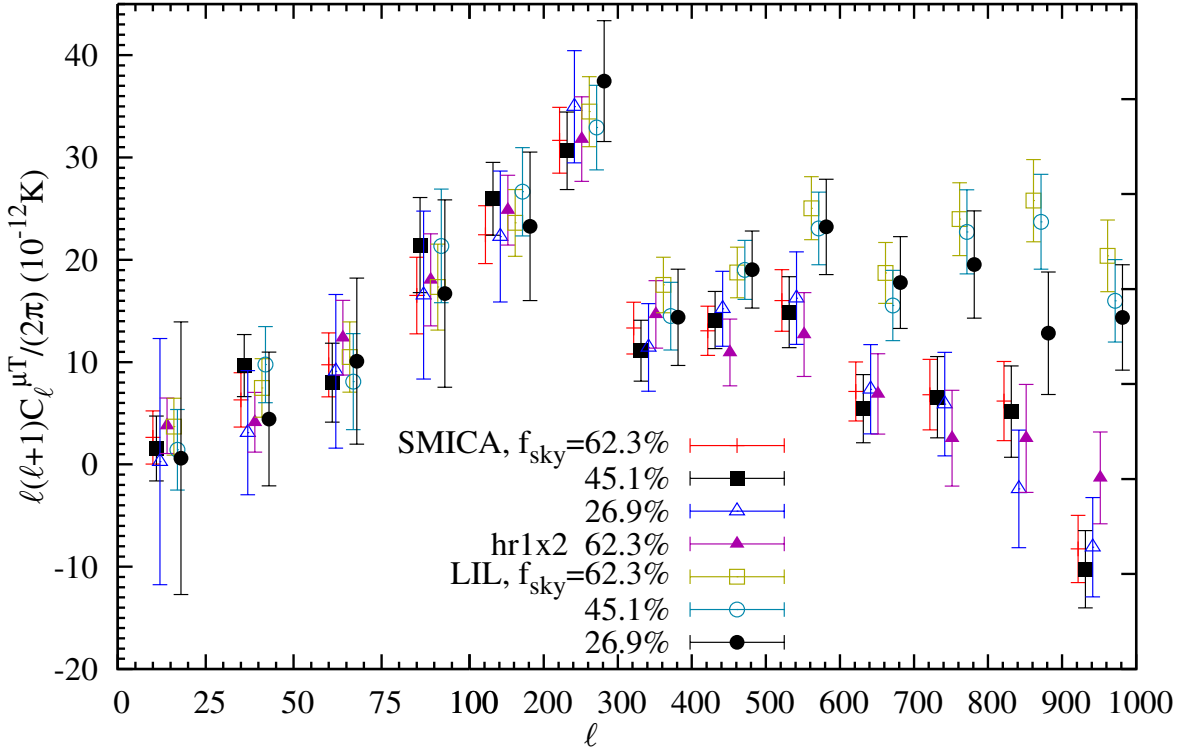}}
\caption{\label{Fig:clmT}Cross power spectrum of the $\mu$-type distortion
  fluctuations with CMB temperature anisotropies. The CMB temperature
  anisotropy maps released by Planck collaboration (SMICA) \cite{planckcmb2015} as well
  as those created by LIL in \cite{k2015} have been used. The effects of
  the effective beams of the respective maps  
  and masks have been corrected for. The points for different masks and map
  combinations have been
offset on the x-axes from the center of the bins to make them
distinguishable. At high $\ell$ we use $\Delta \ell=100$ to make the plot
less cluttered.}
\end{figure}
The cross spectrum of $\mu$ with primary CMB temperature anisotropies $T$
is given by
\begin{align}
\langle a_{\ell m}^ia_{\ell m}^j\rangle =C_{\ell}^{\mu T}+ \langle n_{\ell
  m}^i n_{\ell m}^j\rangle,
\end{align}
where $i$ is the $\mu$-distortion map, either one of the half-ring maps or
the full channel map. Similarly $j$ is one of the CMB temperature
anisotropy maps and $n_{\ell m}^{i,j}$ is the noise in the respective map.
As before we can choose the full mission maps and estimate and subtract the
noise cross power spectrum estimated from the HRHD
maps to get $=C_{\ell}^{\mu T}$ or we can use
one of the  half-ring $\mu$-distortion maps and the other half-ring map for
$T$ in which case the cross noise term would vanish. We again expect the
HRHD maps to give slightly biased estimates of the noise because of the non-linearity
of our component separation algorithm. However on large scales, the
cross-spectrum is dominated not by noise but the CMB anisotropies and we
expect the effect of the noise to be sub-dominant. This can be seen in
Fig. \ref{Fig:clmT} where we have plotted the cross power spectra between
our $\mu$-map and the SMICA (Spectral Matching Independent Component Analysis) CMB map released by the Planck collaboration
\cite{planckcmb2015} as well as between our $\mu$-map and the CMB map produced by LIL
when fitting a $y$+CMB+dust model \cite{k2015}, so that we have a CMB map where the
$y$-type distortion signal is explicitly removed. All power spectra, except
one labeled 'hr1x2', are
calculated from the full channel maps and the noise power calculated from the
HRHD maps has been subtracted. The power spectrum labeled 'hr1x2' is
calculated using the first half-ring $\mu$-map and second half-ring SMICA
CMB map. The cross-power spectrum $C_{\ell}^{\mu T}$ is clearly dominated
by the residual primary CMB anisotropies. This is the same level of
the signal which is also seen if we cross correlate NILC $y$-distortion
map \cite{plancky} and the SMICA CMB map. Thus it seems that for Planck we
are limited not by the sensitivity but by the number of channels, the
latter limiting our ability to separate the spectral distortions from the
primary CMB. The SMICA map is much less noisier and less contaminated on
large scales because they use more channels than the 4 channels used by LIL
and we use the cross spectrum with SMICA for $62.3\%$ of sky using full
channel data to get the limit
\begin{align}
\ell(\ell+1)C_{\ell}^{\mu T}/(2\pi)|_{\ell=2-26}<(2.6\pm
2.6) \times 10^{-12} ~{\rm K}
\end{align}

We note that on the large scales all points are  consistent while on small
scales there is positive correlation between the LIL-CMB and $\mu$ maps
but negative correlation between the SMICA CMB and $\mu$ maps. This is
just the  result of different component separation strategies, in
particular the use of $\ell$-dependent component separation in SMICA.

\section{Theory of fluctuations of $\mu$-type distortions from Silk damping}
The possibility of constraining the primordial non-Gaussianity using the
fluctuations of the $\mu$-type distortions was first presented by
\cite{pajer2012}. Since then there has been considerable work exploring
this possibility in detail for many different sources of primordial
non-Gaussianity 
including inflationary models with primordial magnetic fields and isocurvature perturbations
 \cite{ganc2012,biagetti,miyamoto2014,ganc2014,ota2015,edck2015,naruko2015,shiraishi2015}. We will follow a new approach to the $\mu$-distortion fluctuations from Silk damping motivated by the interpretation of the Silk damping of acoustic modes as mixing of blackbodies \cite{cks2012,ksc2012b}. This approach allows for a simple but precise treatment of the $\mu$-distortion fluctuations and obviates the need for 'averaging over an oscillation' and window functions to localize the dissipated energy. We will work in conformal Newtonian gauge  \cite[e.g.][]{mabert95} following the conventions for the metric perturbations as defined in \cite{dod}.

In the primordial plasma before recombination, photons and baryons are
tightly coupled together through Thomson scattering. The mean free path of
photons is very small but they still diffuse to lengths much larger than
the mean free path by doing a random walk in a sea of electrons. The
primordial energy density fluctuations therefore are washed out on this
diffusion scale, denoted with comoving wavenumber $\kD(\eta)$, and this process is known as Silk damping \cite{silk}. Since the primordial density fluctuations locally
had a Planck spectrum because of the thermalization at $z>2\times 10^6$
\cite{sz1970,dd1982,ks2012}, the photon diffusion locally just mixes the photons
from different blackbodies. The immediate result of the mixing of
blackbodies is $y$-type distortion \cite{zis1972} which rapidly relaxes to equilibrium
Bose-Einstein spectrum at $z\gtrsim z_{\mu}= 5\times 10^4$
\cite{sz1970,is1975b,is1975,bdd1991,cs2011,ks2012,ks2012b}. At smaller redshifts
the spectrum remains a $y$-distortion spectrum. The change from $y$-type
distortion to $\mu$-type distortion is in reality not abrupt but gradual
through an epoch of intermediate-type (or $i$-type) distortions
\cite{is1975b,bdd1991,cs2011,ks2012b} where the initial $y$-type spectrum
has relaxed only partially towards the equilibrium Bose-Einstein distribution. We have chosen
$z_{\mu}$ to be near the center of the intermediate-type region and will
ignore the $i$-type distortions for simplicity as we do not expect them to
affect the constraints from Planck. The dissipation of primordial
fluctuations not only adds energy but also the photons to the monopole component
of the radiation. The $\mu$-distortion created is then given by the total
energy dissipated minus the correction from adding photons \cite{sz1970,ks2012}
\begin{align}
\mu = 1.4\left(\frac{\Delta E}{E}-\frac{4}{3}\frac{\Delta N}{N}\right),\label{Eq:muen}
\end{align}
where $\Delta E/{E}$ is the fractional change in the photon energy density
and $\Delta N/{N}$ is the fractional change in the photon number
density. This simple equation is just the result of fitting two parameters
of the Bose-Einstein spectrum, temperature and chemical potential, to two
constraints, the given energy density and number density of photons.

For Silk damping the above equation implies that \cite{cks2012,ksc2012b}
\begin{align}
\frac{\id \mu(\bx)}{\id \eta}=-2.8\frac{\id}{\id \eta}\left\langle
  \left(\frac{\Delta T(\bx,\hbn)}{T}\right)^2\right\rangle, \label{Eq:mux}
\end{align}
where $\bx$ is the comoving position coordinates, $\eta$ is conformal time,
and $\frac{\Delta T(\bx,\hbn)}{T}$ is the  CMB anisotropy  at
position $\bx$ in direction $\hbn$. We will use bold symbols for vectors
with $\hat{}$ denoting a unit vector and the same symbol in normal font
indicating the amplitude of the vector. Note that $\bx$ and $\hbn$ are
independent variables (i.e. independent of each other) and describe the 5 of the coordinates in 6-dimensional
position-momentum phase space. All terms are functions of $\eta$ which we
will suppress for simplicity unless necessary for clarity. Note that the second term in
Eq. \ref{Eq:muen} has resulted only in $1/3$ of the energy density in the
first term going to the $\mu$-type distortion \cite{cks2012,ksc2012b}. The angular brackets represent an
appropriate average that we will describe now precisely. This average,
according to our formalism,  should lead to mixing of blackbodies
locally. 

The  acoustic dissipation is represented in the Boltzmann hierarchy by the
non-vanishing of the photon quadrupole, even in the tight coupling limit
\cite[see e.g.][]{dod,weinberg}. This can be understood as follows. The
photons diffusing from different regions/direction to a particular position
coordinate  have different temperatures. The electrons therefore see
non-zero anisotropies and Compton scattering isotropizes the radiation
field seen by the electron suppressing the anisotropies. Higher the multipole order more strongly it is
suppressed. The largest anisotropy therefore resides in the quadrupole and
its dissipation in hydrodynamic terms can be understood as the action of
shear viscosity \cite{ksc2012b}. The
mixing of blackbodies by Compton scattering is therefore locally just an
angular average and this is precisely what the angular brackets in Eq. \ref{Eq:mux}
represent. Taking the Fourier transform of Eq. \ref{Eq:mux} and performing
the angular average we get,
\begin{align}
\frac{\id \mu(\bk)}{\id \eta}=-2.8\frac{\id}{\id
  \eta}\int\frac{\id^3\bkp}{(2\pi)^3}\int  \frac{\id\hbn}{4\pi}~ \Theta(\bkp,\hbn)\Theta(\bk-\bkp,\hbn)\label{Eq:muk}
\end{align}
where $\Theta(\bk,\hbn)$ is the Fourier transform of temperature anisotropy
and contains a stochastic part coming from the initial curvature
perturbations , $\mR(\bk)$, and a deterministic transfer function which we will
denote in harmonic space by $\Theta_{\ell}(k)$. The product of anisotropies
in real space gives the convolution in Fourier space. We decompose the
anisotropies into spherical harmonic coefficients, utilizing the fact that
at linear order the transfer functions depend only on the scalar product
$\hbk.\hbn$.
\begin{align} 
\Theta(\bk,\hbn)\equiv\sum_{\ell}(-i)^{\ell}(2\ell+1)\mP_{\ell}(\hbn.\hbk)\Theta_{\ell}(k)\mR(\bk),
\end{align}
where $\mP_{\ell}$ are the Legendre polynomials. Substituting in
Eq. \ref{Eq:muk} and integrating over $\hbn$ we get
\begin{align}
\frac{\id \mu(\bk)}{\id \eta}=-2.8\frac{\id}{\id
  \eta}\int\frac{\id^3k'}{(2\pi)^3}\sum_{\ell m}(-1)^{\ell}(4\pi)Y_{\ell
  m}^{\ast}(\hbkp)Y_{\ell m}(\widehat{\bk-\bkp})\Theta_{\ell}(k')\Theta_{\ell}(|\bk-\bkp|)\mR(\bkp)\mR(\bk-\bkp),\label{Eq:dmudeta}
\end{align}
where we have used the addition theorem for spherical harmonics,
\begin{align}
\mP_{\ell}(\hbk.\hbn)=\frac{4\pi}{2\ell+1}\sum_m Y_{\ell
  m}^{\ast}(\hbk)Y_{\ell m}(\hbn)
\end{align}
Since we are explicitly doing the angular average of CMB anisotropies
locally at position $\bx$, our energy dissipation is localized by
construction and we do not need a window function to localize the
dissipated energy. 

In the tight coupling limit almost all of the energy is in the monopole and
dipole. The tight coupling solution for the linear CMB transfer functions
is \cite{sz1970c,Peebles1970,dzs1978,seljak1994,hs1995,mabert95}
\begin{align}
\Theta_0&\approx
A^i(-3-0.4\Rnu)\cos(k\rs)e^{-k^2/\kD^2}\nonumber\\
\Theta_1&\approx
A^i\frac{(-3-0.4\Rnu)}{\sqrt{3}}\sin(k\rs)e^{-k^2/\kD^2}\\
\Theta_{\ell}&\approx 0 ~~{\rm if}~~\ell\ge 2\nonumber\\
A^i&\equiv\frac{0.5}{0.4\Rnu+1.5}=0.3\nonumber
\end{align}
where $\rs(\eta)=\int_0^{\eta}\id \eta c_{\rm s}$ is the sound horizon,
$c_{\rm s}$ is the sound speed of baryon-photon fluid which at $z\gtrsim
5\times 10^4$ is well approximated by $c_{\rm s}\approx 1/\sqrt{3}$, $\kD(\eta)$
  is the damping wave number  \cite{silk,Peebles1970,kaiser}. These
  analytic solutions assume that the two metric perturbations $\phi$ and
  $\psi$ have the same transfer function in the radiation dominated era,
\begin{align}
\left(\phi,\psi\right)=3\left(\phi^i,\psi^i\right)\left(\frac{\sin(k\eta/\sqrt{3})-(k\eta/\sqrt{3})\cos(k\eta/\sqrt{3})}{(k\eta/\sqrt{3})^3}\right),
\end{align}
where ${}^i$ indicates the initial perturbation on super horizon scales in the
radiation dominated epoch but after neutrino decoupling. These solutions take into
account the effect of neutrino  anisotropic stress in the initial conditions,
$\phi^i=-(1+0.4R_{\nu})\psi^i, \Theta_0^i=-\psi^i/2,
\psi^i=\mR/(0.4\Rnu+1.5)$, where
$\Rnu=\rho_{\nu}/(\rho_{\gamma}+\rho_{\nu})\approx 0.409$ is the fraction
of neutrino energy density with respect to the total (photons+neutrino)
energy density in the radiation dominated epoch after the electron-positron
annihilation. At the level of accuracy needed for us, we can set
$3+0.4\Rnu\approx 3$.

Substituting the tight coupling solutions into Eq. \ref{Eq:dmudeta} and taking  the
ensemble  average of the stochastic part, $\mR$ terms, we recover the
results of \cite{cks2012,ksc2012b}. It is interesting to note that the sum
of monopole and dipole terms gives a term $\propto
\cos^2(k\rs)+\sin^2(k\rs)=1$, i.e. the total energy in the acoustic
oscillations adds up to a non-oscillating constant. This is as expected
since the sound waves just represent the oscillation of internal energy
($\Theta_0$) and kinetic energy ($\Theta_1$) into each other while the
total energy of the sound wave must be conserved in the lab frame (at rest
with respect to the bulk of the fluid). The
energy (and therefore the above interpretation) is of course a gauge or
reference frame dependent quantity and we may interpret the above solutions
differently in a different gauge \cite[e.g.][]{pz2012}. In our formalism
 there is no need to do the average over an oscillation period of
the sound wave as the total contribution to the $\mu$-type distortion does
not oscillate.

Since we are interested in fluctuations of $\mu$, we will defer the statistical or ensemble average
of the stochastic component, $\mR$, until the end when we calculate the
angular power spectra. At high redshifts, $z\gtrsim 2\times 10^6$, the
$\mu$-type distortions are suppressed as the photon creation by double
Compton scattering and bremsstrahlung together with comptonization is able
 relax the spectrum to Planck spectrum, $\mu\longrightarrow 0$. We can take
 this into account by multiplying the $\mu$ distortion by the blackbody
 visibility function \cite{sz1970,dd1982,ks2012}, $\mG(\eta)$. Integration
  up to $z=z_{\mu}\approx 5\times 10^4$ then gives us the $\mu$-type
  distortions at $z_{\mu}$, or the corresponding conformal time $\eta_{\mu}$,
\begin{align}
\mu(\bk,\eta_{\mu})=& -2.8\int_0^{\eta_{\mu}}\id \eta~\mG(\eta)\frac{\id}{\id
  \eta}\int\frac{\id^3k'}{(2\pi)^3}\sum_{\ell m}(-1)^{\ell}(4\pi)Y_{\ell
  m}^{\ast}(\hbkp)Y_{\ell m}(\widehat{\bk-\bkp})\nonumber\\
&\Theta_{\ell}(k',\eta)\Theta_{\ell}(|\bk-\bkp|,\eta)\mR(\bkp)\mR(\bk-\bkp).
\end{align}
Precise expressions for $\mG$ are given in \cite{ks2012} but it is well
approximated by the simple expression, sufficient for our purpose, $\mG\approx e^{-(z/z_{\rm
    dC})^{5/2}}$, where $z_{\rm dC}\approx 2\times 10^6$.

We will be interested in the squeezed limit, where we want to calculate the
fluctuations that are much larger than the dissipation scale. The
dissipation scale here is given by $k'$ and the fluctuation scale by
$k$. We can therefore take the limit $k\longrightarrow 0 $ in the
non-stochastic terms to simplify the
above expression,
\begin{align}
\mu(\bk,\eta_{\mu})&\overset{k\rightarrow 0}{\approx}  -2.8\int_0^{\eta_{\mu}}\id \eta~\mG(\eta)\frac{\id}{\id
  \eta}\int\frac{\id^3k'}{(2\pi)^3}\sum_{\ell}(2\ell+1)
\Theta_{\ell}(k',\eta)\Theta_{\ell}(|\bk-\bkp|,\eta)\mR(\bkp)\mR(\bk-\bkp)\nonumber\\
&\approx  -2.53\int_0^{\eta_{\mu}}\id \eta~\mG(\eta)\frac{\id}{\id
  \eta}\int\frac{\id^3k'}{(2\pi)^3}
e^{-2k'^2/\kD^2}\mR(\bkp)\mR(\bk-\bkp),
\end{align}
where we have used the tight coupling solution in the last line where the
oscillating part evaluates to a constant in the limit $k\rightarrow 0$.
The time dependent part is now very simple and is trivially integrated if
we approximate $\mG$ by a step function that is 
unity for redshifts in the range  $5\times 10^4 \equiv z_{\mu}\le z\le \zpl \equiv 2\times 10^6$ and zero
otherwise, yielding a final very simple expression
for the $\mu$-distortion fluctuations
\begin{align}
\mu(\bk,\eta_{\mu}) \approx 2.53\int\frac{\id^3k'}{(2\pi)^3}
\left[e^{-2k'^2/\kD(\zpl)^2}-e^{-2k'^2/\kD(z_{\mu})^2}\right]\mR(\bkp)\mR(\bk-\bkp).
\end{align}

The fluctuations in $\mu$ represent fluctuations in the energy density of
photons and these energy density fluctuations would oscillate in the same
way as the primary CMB fluctuations. The $\mu$ parameter that we have
defined is however gauge invariant, similar to the dimensionless frequency
$x=\nu/T$ and does not undergo oscillations. To be more specific, the adiabatic part of the radiation
transfer function has no effect on fluctuations of $\mu$. The $\mu$
fluctuations are therefore frozen in at $z=z_{\mu}$. The dissipative or
Silk damping part of the radiation transfer function would affect the
$\mu$-fluctuations and $\mu$-type fluctuations would also dissipate on the
diffusion scale which keeps on increasing until the recombination. The
radiative transfer of the $\mu$-type fluctuations is therefore very simple, it is
modified by Silk damping in the same way as the primary CMB anisotropies
until recombination 
and after recombination they just free stream to us. We can take into
account the Silk damping very simply to get the $\mu$-distortion fluctuations at
the last scattering surface (LSS), $z\approx 1100$,
\begin{align}
\mu(\bk,\eta_{LSS})=\mu(\bk,\eta_{\mu})e^{-k^2/\kD(\eta_{\rm LSS})^2},
\end{align}
where $\kD(\eta_{\rm LSS})$ is the damping wavenumber at the last scattering
surface which we can evaluate numerically for standard recombination
history from the well known expression
\cite{silk,Peebles1970,kaiser,dod,weinberg}. It is interesting to note that
the dissipation of $\mu$ fluctuations will result in mixing of
Bose-Einstein spectra giving a $y^{\rm BE}$-type distortion  different from the $y$-type
distortion coming from the mixing of Planck spectra. We can of course
trivially calculate this distortion using the same formalism used for
mixing of blackbodies. For fluctuations $\delta_{\mu}$ around a mean
$\bar{\mu}$, we get after mixing a distortion $n_{y^{\rm BE}}$ given by
\begin{align}
n_{y^{\rm
    BE}}=\frac{1}{2}\frac{\bar{\mu}^2\left\langle\delta_{{\mu}}^2\right\rangle e^{x+\bar{\mu}}}{\left(e^{x+\bar{\mu}}-1\right)^2}\left(\frac{e^{x+\bar{\mu}}+1}{e^{x+\bar{\mu}}-1}\right),
\end{align}
where we have expanded in Taylor series the spectrum
$1/(e^{x+\bar{\mu}(1+\delta_{\mu})}-1)$. The zeroth order term is just the
mean Bose-Einstein spectrum, the first order term vanishes since $\left\langle
\delta_{\mu}\right\rangle =0$ by definition and the second order term gives
the lowest order distortion. 
 This distortion would be an effect that is 4th
order in perturbations, since $\mu$ is already 2nd order and we will not pursue this
interesting possibility here. Note that the temperature of the Bose-Einstein
spectrum would also vary along with the $\mu$ and will give the usual
$y$-type distortion at lowest order.

 We should note that for modes outside the
horizon at the time of recombination, in the Sachs-Wolfe (SW) limit
\cite{sw1967}, the effect of Silk damping is negligible and we can ignore
it,
\begin{align}
\mu^{\rm SW}(\bk,\eta_{\rm LSS})\approx \mu(\bk,\eta_{\mu})
\end{align}

After the recombination the $\mu$-type fluctuations just free stream to us,
with a small damping at reionization that we will ignore here, and the
equation describing it is just the free streaming Boltzmann equation,
\begin{align}
\frac{\partial \mu(\bk,\eta)}{\partial \eta}+ik\hbk.\hbn\mu(\bk,\eta)=0
\end{align}
Integrating from $\eta_{\rm LSS}$ to today $\eta_0$ we get,
\begin{align}
\mu(\bk,\eta_0)&=\mu(\bk,\eta_{\rm LSS})e^{-ik\hbk.\hbn(\eta_0-\eta_{\rm
    LSS})}\nonumber\\
&=\mu(\bk,\eta_{\rm LSS})4\pi\sum_{\ell'}(-i)^{\ell'}j_{\ell'}\left(k(\eta_0-\eta_{\rm
  LSS})\right)\sum_{m'}Y_{\ell' m'}(\hbn)Y_{\ell' m'}^{\ast}(\hbk),
\end{align}
where we have used the expansion of the exponential function in spherical
harmonics \cite{varsh}.
We can now Fourier transform it back to real space and decompose it into
spherical harmonic coefficients giving,
\begin{align}
\mu_{\ell m}(\bx)&=4\pi(-i)^{\ell}\int \frac{\id^3 k}{(2\pi)^3}e^{i\bk.\bx}j_{\ell}\left(k(\eta_0-\eta_{\rm
  LSS})\right)Y_{\ell m}^{\ast}(\hbk)\mu(\bk,\eta_{\rm LSS}) \nonumber\\
&\approx 10.12\pi(-i)^{\ell}\int \frac{\id^3 k}{(2\pi)^3}\frac{\id^3 k'}{(2\pi)^3}e^{i\bk.\bx}j_{\ell}\left(k(\eta_0-\eta_{\rm
  LSS})\right)Y_{\ell m}^{\ast}(\hbk)\nonumber\\
&\times \left[e^{-2k'^2/\kD(\zpl)^2}-e^{-2k'^2/\kD(z_{\mu})^2}\right]e^{-k^2/\kD(\eta_{\rm LSS})^2}\mR(\bkp)\mR(\bk-\bkp),
\end{align}
where  $\bx$ is the position of the observer. This is our final expression
for $\mu$-distortion anisotropies which agrees with the corresponding
expression in \cite{pajer2012} apart from the  trigonometric and the window
functions which are not needed in our formalism. The damping factor for $\mu$-type fluctuations, $e^{-k^2/\kD(\eta_{\rm LSS})^2}$, can be
neglected in the Sachs-Wolfe limit but must be included when using modes
inside the horizon during recombination ($\ell \gtrsim 100$). For a statistically
homogeneous Universe, such as ours is assumed to be, the statistically
averaged quantities would not depend on the position of the observer. The
dependence on $\bx$ would therefore disappear once we take the statistical
average, similar to the calculations of the primary CMB anisotropies. To be specific, once
we do the statistical average of two quantities with wavenumbers $\bk_1$ and
$\bk_2$ as arguments, such as for the calculation of the power spectrum, we will get the factor
$\delta_D^3(\bk_1+\bk_2)e^{i(\bk_1+\bk_2).\bx}$ making the argument of the
exponent and hence the dependence on the position of the observer vanish
\cite[see e.g.][]{dod}. The free streaming solution therefore just projects
the 3-d fluctuations at the last scattering surface onto a sphere in the
spherical harmonic basis. This completes our derivation of the
evolution of $\mu$-distortion fluctuations and the formulae are completely
analogous to those of primary CMB anisotropies if we make  the instant recombination
approximation and free stream the CMB anisotropies from the LSS to today,
\begin{align}
a_{\ell m}&= 4\pi(-i)^{\ell}\int \frac{\id^3
  k}{(2\pi)^3}e^{i\bk.\bx}Y_{\ell
  m}^{\ast}(\hbk)\Theta_{\ell}(k,\eta_0)\mR(\bk)\nonumber\\
&\approx 4\pi(-i)^{\ell}\int \frac{\id^3
  k}{(2\pi)^3}e^{i\bk.\bx}j_{\ell}\left(k(\eta_0-\eta_{\rm
  LSS})\right)Y_{\ell
  m}^{\ast}(\hbk)S(k,\eta_{\rm LSS})\mR(\bk),
\end{align}
where $S(k,\eta_{\rm LSS})$ is the source term that reduces in the
Sachs-Wolfe approximation to 
\begin{align}
S^{\rm SW}(k,\eta_{\rm LSS})=\Theta_0(k,\eta_{\rm LSS})+\psi(k,\eta_{\rm
  LSS})\approx \frac{\psi(k,\eta_{\rm
  LSS})}{3}\approx \frac{1}{5}
\end{align}

\subsection{The auto and cross power spectrum of $\mu$-distortion
  fluctuations}
We will assume the following local model for the primordial curvature
perturbations:
\begin{align}
\mR(\bx)=\mR_G(\bx)+\frac{3}{5}\fnl\mR_G(\bx)^2,
\end{align}
where $\mR_G$ is a Gaussian random field and $\fnl$ is the non-linear
parameter. With this model the cross-power spectrum between $\mu$ and CMB
temperature $C_{\ell}^{\mu T}\propto \fnl$ and the auto power spectrum of
$\mu$, $C_{\ell}^{\mu \mu}\propto 9/25\fnl^2\equiv \taunl$.

We will denote by the angular brackets the statistical ensemble average  from
now on. For the cross power spectrum we get
\begin{align}
\left\langle a_{\ell m}\mu_{\ell' m'}\right\rangle = (-1)^m\delta_{\ell \ell'}\delta_{m -m'}C_{\ell}^{\mu
  T},
\end{align}
where $\delta$ represents the Kronecker delta function and 
\begin{align}
C_{\ell}^{\mu T}=&\frac{10.12}{\pi^3}\frac{3}{5}\fnl \int \id k k^2
j_{\ell}^2\left(k(\eta_0-\eta_{\rm LSS})\right)S(k,\eta_{\rm
  LSS})P_{\mR_G}(k)\nonumber\\
&\times  \int\id k' k'^2
\left[e^{-2k'^2/\kD(\zpl)^2}-e^{-2k'^2/\kD(z_{\mu})^2}\right] P_{\mR_G}(k'),
\end{align}
where we have used, taking the limit $k\ll k'$,
\begin{align}
\left\langle \mR(\bk')\mR(\bk-\bkp)\mR(\bk_2)\right\rangle \approx \frac{6}{5}(2\pi)^3\fnl\delta_D^3(\bk+\bk_2)\left[P_{\mR_G}(k')^2+2P_{\mR_G}(k')P_{\mR_G}(k)\right].
\end{align}
The first term in the square brackets gives a term of the form $\int \id k
k^2 j_{\ell}(k x)$ independent of the CMB  anisotropies which is formally infinite but in the squeezed limit,
$k\rightarrow 0$, should vanish 
 and we discard it  including
only the second term in expression for the $C_{\ell}^{\mu T}$. The
integrals are easy to do for a spectrum of form
$P_{\mR_G}=A_{\mR_G}(2\pi^2/k^3)(k/k_0)^{\nS-1}$ giving us the final
expression,
\begin{align}
C_{\ell}^{\mu T}\approx\frac{4.86\pi \fnl
  A_{\mR_G}^2(\eta_0-\eta_{\rm LSS})^{1-\nS}}{k_0^{2\nS-2}}f_{\ell}(\nS),
\end{align}
where,
\begin{align}
f_{\ell}(\nS\ne 1)&=\frac{\sqrt{\pi}}{4 }\frac{\Gamma\left(\frac{3-\nS}{2}\right)\Gamma\left(\frac{\nS-1+2\ell}{2}\right)}{\Gamma\left(\frac{4-\nS}{2}\right)\Gamma\left(\frac{5+2\ell-\nS}{2}\right)}\left[2^{-(\nS+1)/2}\Gamma\left(\frac{\nS-1}{2}\right)\left(\kD(\zpl)^{\nS-1}-\kD(z_{\mu})^{\nS-1}\right)\right]\nonumber\\
f_{\ell}(\nS=1)&=\frac{1}{2\ell(\ell+1)}\log\left(\frac{\kD(\zpl)}{\kD(z_{\mu})}\right)\label{Eq:fl},
\end{align}
where $\Gamma$ is the gamma function.

For the auto power spectrum we get,
\begin{align}
C_{\ell}^{\mu\mu}=&\taunl\frac{10.12^2}{2\pi^5}\int \id k ~k^2
j_{\ell}\left(k(\eta_0-\eta_{\rm LSS})\right)^2P_{\mR_G}(k)e^{-2k^2/\kD(z_{\rm LSS})^2}\nonumber\\
&\times \left(\int\id k' k'^2
\left[e^{-2k'^2/\kD(\zpl)^2}-e^{-2k'^2/\kD(z_{\mu})^2}\right]
P_{\mR_G}(k')\right)^2,
\end{align}
where we have used
\begin{align}
\left\langle
  \mR(\bkp_1)\mR(\bk_1-\bkp_1)\mR(\bkp_2)\mR(\bk_2-\bkp_2)\right\rangle
=\frac{36}{25}\fnl^2(2\pi)^3\delta_D^3(\bk_1+\bk_2)\left[4P_{\mR_G}(k_1)P_{\mR_G}(k'_1)P_{\mR_G}(k'_2)+f(k'_1,k'_2)\right],
\end{align}
where $f(k'_1,k'_2)$ represents permutations which are independent of
$k_1,k_2$ in the squeezed limit, $k_1,k_2\rightarrow 0$ and would again result in
terms in $C_{\ell}$ of form $\int \id k
k^2 j_{\ell}(k x)$ which should vanish in the limit $k\rightarrow 0$. We
have also defined $\taunl\equiv 9/25 \fnl^2$.
The integrals are again analytically solvable on large scales for power law initial
curvature spectrum giving
\begin{align}
C_{\ell}^{\mu\mu}=&4\pi\taunl\frac{10.12^2A_{\mR_G}^3}{k_0^{3(\nS-1)}}(\eta_0-\eta_{\rm LSS})^{1-\nS}g_{\ell}(\nS),
\end{align}
where
\begin{align}
g_{\ell}(\nS\ne 1)
&=\frac{\sqrt{\pi}}{4}\frac{\Gamma\left(\frac{3-\nS}{2}\right)\Gamma\left(\frac{\nS-1+2\ell}{2}\right)}{\Gamma\left(\frac{4-\nS}{2}\right)\Gamma\left(\frac{5+2\ell-\nS}{2}\right)}\left[2^{-(\nS+1)/2}\Gamma\left(\frac{\nS-1}{2}\right)\left(\kD(\zpl)^{\nS-1}-\kD(z_{\mu})^{\nS-1}\right)\right]^2\nonumber\\
g_{\ell}(\nS= 1) &=\frac{1}{2\ell(\ell+1)}\left[\log\left(\frac{\kD(\zpl)}{\kD(z_{\mu})}\right)\right]^2\label{Eq:gl}
\end{align}
We should point out that we can choose different parameter values for the
primordial power spectrum on large and small scales in the above
expressions. One factor of $A_{\mR_G}/k_0^{\nS-1}\times $ the
term in square brackets in Eq. \ref{Eq:fl} in the expression for
$C_{\ell}^{\mu T}$ corresponds to the small scale
perturbations and it is possible to choose different $A_{\mR_G},k_0,\nS$ for
these factors to take into account, for example, running of the spectral
index or a break in the power spectrum. The same is true for
$(A_{\mR_G}/k_0^{\nS-1})^2\times$ the term in the square brackets in
Eq. \ref{Eq:gl} for the $C_{\ell}^{\mu\mu}$.

\subsection{Constraints on primordial non-Gaussianity from $\mu$-distortion
fluctuations}
Using the $\Lambda$CDM cosmology parameters \cite{wmap,planck2015},
$\nS=0.965, A_{\mR_G}=2.1\times 10^{-9},\kD(z_{\mu})=46~{\rm
  Mpc}^{-1},\kD(z_{\mu})=1.1\times 10^4~{\rm Mpc}^{-1},k_0=0.05~{\rm Mpc^{-1}},
\eta_0-\eta_{\rm LSS}=13.9 ~{\rm Gpc}$, we get on large scales
\begin{align}
\frac{\ell(\ell+1)}{2\pi}C_{\ell}^{\mu T}\approx 2.4\times
10^{-17}\fnl\nonumber\\
\frac{\ell(\ell+1)}{2\pi}C_{\ell}^{\mu \mu}\approx 1.7\times
10^{-23}\taunl,
\end{align}
where we have evaluated the expressions at $\ell=13$, the center of our
$\Delta \ell=25$ bin in Planck spectra,  and the dependence on
$\ell$ on large scales is very mild because $\nS$ is very close to $1$. If
we choose $\nS=1$, with the other parameters the same, we get slightly
larger signal as expected,
\begin{align}
\frac{\ell(\ell+1)}{2\pi}C_{\ell}^{\mu T}(\nS=1)= 2.9\times
10^{-17}\fnl~{\rm K}\nonumber\\
\frac{\ell(\ell+1)}{2\pi}C_{\ell}^{\mu \mu}(\nS=1)= 2.8\times
10^{-23}\taunl
\end{align}

Finally we get the constraints on the non-linear parameters using our
measurements of the $\mu$-fluctuations,
\begin{align}
\fnl&= \frac{\ell(\ell+1)}{2\pi}\frac{C_{\ell}^{\mu T}}{2.4\times
  10^{-17}}\nonumber\\
&<  10^5\\
\taunl&=\frac{\ell(\ell+1)}{2\pi}\frac{C_{\ell}^{\mu \mu}}{1.7\times
  10^{-23}}\nonumber\\
&<1.4\times 10^{11}
\end{align}

\section{Conclusions}
We have constructed all sky maps of $\mu$-distortion fluctuations from
the publicly available Planck HFI data. Our maps and masks are made
publicly available at \url{http://www.mpa-garching.mpg.de/~khatri/muresults/}. Since we have limited number of
channels in Planck, especially channels not dominated completely by
foregrounds, it is not possible to completely separate the $\mu$-type and
$y$-type distortions and our $\mu$-type distortions maps are dominated by
the $y$-distortion contamination. Nevertheless, it is possible to put
interesting upper limits {on the $\mu$ anisotropy power spectrum
  and cross spectrum of $\mu$ with CMB temperature anisotropies. We have
  presented measured angular auto and cross power spectra for $\ell
  \lesssim 1000$ which should be taken as the upper limits.}

{Our observational constraints on $C_{\ell}^{\mu\mu}$ and $C_{\ell}^{\mu T}$
allows us to constrain new physics which may be responsible for spatially
varying energy release in the early Universe, $5\times 10^4\lesssim
z\lesssim 2\times 10^6$. We apply our results to constrain  the primordial non-Gaussianity
 for extremely squeezed configurations}, $k_{\rm S}/k_{\rm L}\approx 5\times
 10^4- 10^7$, where very few constraints exist at present. Our
 constraints are much stronger, for the same scales, than the recent constraints obtained by
 \cite{naruko2015} using only the CMB temperature anisotropies. The only
 other comparable constraints, to our knowledge, come from the considerations of formation of
 primordial black holes from collapse of primordial fluctuations in the
 presence of large non-Gaussianity \cite{jgm2009,bcg2012}. Our constraints
 for the non-linear parameters are $\fnl<10^5, \taunl<1.4\times
 10^{11}$ implying that the non-Gaussianity on these extremely small
 scales, is
 smaller than of order unity. {We should note that the mean or
   average distortion $\langle \mu\rangle$ should be of same order of
   magnitude as the rms, $\mu_{\rm rms}$, i.e. a strong non-Gaussianity of
   local type would also increase the average $\mu$-distortion compared to
   the case of no non-Gaussianity, other parameters being the same. Thus the constraints from
   COBE-FIRAS on $\langle \mu\rangle$ also constrain primordial
   non-Gaussianity. Our constraints using the Planck data are however much
   stronger and in particular constrain the non-linear term $\fnl\mR_{G}^2$ to
   be not larger than the linear term $\mR_G$ in the primordial perturbation $\mR$.} With the future experiments, such as Pixie (Primordial Inflation
 Explorer) and LiteBIRD (Lite satellite for the studies of B-mode
 polarization) \cite{litebird},  it will be possible to improve on these limits by many orders
 of magnitude. However, with Planck satellite mission we have already entered a new era of
  CMB spectrum cosmology.

\acknowledgments
We would like to thank Nail Inogamov for discussions on the hydrodynamic
aspects of the sound waves, in particular the conservation of total energy.
This paper used observations obtained with Planck
(\url{http://www.esa.int/Planck}), an ESA science mission with instruments and
contributions directly funded by ESA Member States, NASA, and Canada. We
 also acknowledge  use of the HEALPix software \cite{healpix}
 (\url{http://healpix.sourceforge.net}).
 This research has made use of "Aladin sky atlas" developed at CDS,
 Strasbourg Observatory, France \cite{aladin}. This research has also made use of the SIMBAD database, operated at CDS, Strasbourg, France.
RS acknowledges partial support by grant No. 14-22-00271 from the Russian Scientific Foundation.

\bibliographystyle{unsrtads}
\bibliography{muconstraints}
\end{document}